\documentstyle[preprint,aps]{revtex}

\begin{document}


\draft
\title{Modeling of Dislocation Structures in Materials}
\author{J. M. Rickman}
\address{
Department of Materials Science and Engineering, Lehigh University,
Bethlehem, PA  18015}
\author{Jorge Vi\~nals}
\address{
Supercomputer Computations Research Institute,Florida State University,
Tallahassee, Florida 32306-4052, and Department of Chemical Engineering,
FAMU-FSU College of Engineering, Tallahassee, Florida 32310}

\date{\today}
\maketitle
\begin{abstract}

A phenomenological model of the evolution of an ensemble of interacting
dislocations in an isotropic elastic medium is formulated. The line-defect
microstructure is described in terms of a spatially coarse-grained
order parameter, the dislocation density tensor. The tensor field
satisfies a conservation law that derives from the conservation of Burgers
vector. Dislocation motion is entirely dissipative and
is assumed to be locally driven by the minimization of plastic free energy.
We first outline the
method and resulting equations of motion to linear order in the
dislocation density tensor, obtain
various stationary solutions, and give their geometric interpretation.
The coupling of the dislocation density to an externally imposed stress
field is also addressed,
as well as the impact of the field on the stationary solutions.

\end{abstract}


\section{Introduction}
\label{sec:introduction}

The presence and motion of dislocations in materials is known to affect
many of their mechanical properties, including, for example, resistance to
deformation and plastic response to relatively large stresses
\cite{re:hull92}. Given the impact of dislocation dynamics on material
properties, there have been a number of models advanced attempting to
describe the motion of line defects in
response to both internal and applied stresses, which have substantially
furthered our understanding of these defect-property correlations
\cite{re:hirth68,re:mura63,re:ghonhiem88,re:gulluoglu90}.

The starting point of a large
class of models of dislocation dynamics is the calculation of
direct interactions between a dislocation and a stress field, as given by the
Peach-Koehler force \cite{re:hirth68}. On the one hand
it is possible to construct a Newtonian
model of dislocation motion by combining the law of conservation
of Burgers vector with Newton's second law, a Hookean stress-strain
constitutive law and the differential relation between plastic strain and
dislocation density, as has been done elsewhere. \cite{re:hirth68}  While
this approach has
been very useful in elucidating some important features of dislocation
motion, its inherent limitations to relatively short time scales and small
numbers of dislocations make its generalization to systems with large
numbers of dislocations which evolve over long periods of time somewhat
impractical. Numerical calculations of the dynamical evolution of an
ensemble of pairwise interacting dislocations have been carried out
with Molecular Dynamics methods to study relatively short time scales
of the order of several vibrational periods \cite{re:mura87}.
A different approach to dislocation dynamics is based on purely
dissipative or Langevin dynamics \cite{re:gulluoglu90}. While the
level of description is still microscopic, the motion of each
dislocation line is assumed to be overdamped. This method allows
the study of longer time scales at the price of added phenomenology in
the introduction of friction or damping forces.

While the aforementioned approaches have proved quite fruitful, they
share several drawbacks that limit their usefulness in studying the
evolution of a large ensemble of dislocations over relatively long
periods of time. Among them, we note the long ranged nature of
dislocation-dislocation interactions which must be truncated in
actual calculations, often in some ad-hoc fashion \cite{re:ghonhiem88}.
In addition, the microscopic nature of the description places a
severe restriction to the number of dislocation lines that can be
considered. As a consequence, the scale of the resulting
microstructure is essentially set by the number of dislocations
in a typical simulation cell.
We note, however, that recent computational advances have permitted
studies involving up to $10^{6}$ dislocations \cite{re:gulluoglu89} .

Experimental evidence, on the other hand, suggests that, at
least in some cases, the mechanical response of a material may depend only
weakly on small-scale microstructural details, and more strongly on the
macroscopic density of dislocations. \cite{re:turner82}
Such evidence implies that
an approach based, ${\it a}$ $ {\it priori}$, on the dislocation density
might best capture the important macroscopic features of such systems.
In this paper we follow this approach and
formulate a simplified, linear phenomenological model of dislocation
motion, in terms of the dislocation density tensor. The model properly
incorporates
necessary conservation laws related to the topological structure of the
dislocation lines, as well as the true tensorial character of the
density. By contrast with previous approaches to this problem, and in the
same spirit as irreversible thermodynamics, we employ a
non-deterministic description which focuses on a small number of
macro-variables and treats the remaining degrees of freedom, which are
presumed to evolve on a faster time scale, as an effective heat bath.
This is formally accomplished by applying a coarse-graining procedure to a
system of dislocations in an elastically isotropic medium, and
selecting the dislocation density tensor as the appropriate order
parameter to describe the system at this scale.
Of course,
this approach is formally equivalent to Langevin type models mentioned
earlier: dislocation motion is overdamped because of its interaction
with other microscopic degrees of freedom that act as a \lq\lq
heat bath".
We argue, however, that the overdamped character of dislocation motion
arises only at a semi-macroscopic scale, and that it includes not only
interactions with the \lq\lq heat bath", but also mutual dislocation
interactions within a coarse-graining cell.

Our approach also parallels the study of the equilibrium statistical
mechanics of a system of unbound dislocations undertaken by
Nelson and Toner \cite{re:nelson81}. They investigated the effect
of a collection of dislocations on the broken translational and
rotational symmetries of the crystalline state by examining the
Fourier transform of the dislocation density-density correlation
function at long wavelengths. In so doing, they argued that
dislocations destroy long-ranged translational order, but not
long-ranged orientational order. \cite{re:nelson81}
Central to this calculation is the identification of the
appropriate Boltzmann weight (and therefore the free energy) that determines
the probability of the occurrence of a given fluctuation, and of the
corresponding defect configuration. In their view the relevant dislocation
driving forces can be constructed from a free energy which is a functional
of the dislocation density tensor, with components $\rho_{ij}$, and a
non-singular
strain, with components $e_{ij}^{NS}$, which describes
vibrational degrees of freedom in the crystal. The contribution to
the free energy that depends on $\rho_{ij}$ was chosen equal to
the energy of interaction between dislocations, as derived by
Kosevich \cite{re:kosevich79}.
The logical next step, which we propose here, is to use the minimization
of the singular contribution to the free energy as the
driving force for dislocation motion.

We finally mention complementary earlier work by Brand and Kawasaki
\cite{re:brand84,re:kawasaki84a,re:kawasaki84b}, who derived phase
equations
for generic systems that support topological defects. A phase equation
involves a phase
variable that changes slowly on the scale of the lattice spacing and in
time, and that has a nonzero winding number around a topological defect.
This type of equation represents the long-wavelength distortion of
some ideal basic pattern and has been successfully used in a large
variety of nonequilibrium systems \cite{re:cross93}.
We focus our attention on the case of a crystalline solid and retain the
full dislocation density tensor in our description, since we believe that
this is the natural coarse-grained variable to describe a defected
crystalline solid. Our choice of variables may prove useful in two
different directions. First, the equations that we present below
may provide a starting point for direct numerical calculations of
dislocation motion, a subject of considerable interest.
Second, coupling to other slow variables of interest can be accomplished
quite naturally. These would include, for example, energy
or mass densities. The latter are necessary to describe other irreversible
phenomena such as the effect of mass diffusion on dislocation motion.

This paper is organized as follows. In Sec. \ref{sec:model} we describe the
dynamical
equations governing the evolution of the dislocation density in an isotropic
elastic medium.  In Sec. \ref{sec:stationary} stationary solutions
are identified and given geometrical interpretations, both in systems
with and without externally imposed stresses.  Finally, Sec.
\ref{sec:conclusion} contains a short summary and discussion of future work.

\section{Model Equations}
\label{sec:model}

Consider an isothermal, three-dimensional, linear isotropic elastic medium
which contains a large, finite number of unbound dislocation lines which
are free to move throughout the system.  Suppose further that this medium
is subdivided into a number of hypothetical averaging cells, each of volume
$V_{0}$, such that a relatively large number of dislocation lines thread a
given volume element. Now, if one wishes to describe the defect properties
of this system on a length scale larger than $(V_{0})^{1/3}$, then one can
define a coarse-grained variable, the dislocation density, which
reflects the average dislocation content of a cell. As a dislocation is
characterized by two directions in space, its line direction, ${\hat
\chi}$, and the direction of its Burgers vector, ${\vec b}$, it is expected
that the dislocation density must also carry such information and would,
therefore, be a second-rank tensor. The components of this tensor will be
denoted by $\rho_{ij}$ hereafter.

In modeling the dynamics of the defected system described above at this
coarse-grained
level, it is necessary to include \lq\lq slow" variables, the relaxation of
which takes place on time scales much longer than microscopic times
\cite{re:martin72,re:forster75}. Variables that satisfy global conservation
laws or that arise as a consequence of a spontaneously broken symmetry
satisfy this requirement. Although line defects are not present in
equilibrium
in a three dimensional crystalline solid, we take the view that there
exists a
time scale, slow compared to relevant experimental times, in which the net
Burgers vector in a crystalline solid is conserved. In practice, of course,
dislocations may migrate to the surface of the crystal or to a
grain boundary and disappear.
Conservation of Burgers vector can be expressed in terms of the dislocation
density tensor $\rho_{ij}$ as \cite{re:kosevich79}
\begin{equation}
\label{eq:conserve}
{\partial \rho_{ij} \over \partial t}  + \epsilon_{ilm} {\partial j_{mj}
\over \partial x_{l}}= 0 ,
\end{equation}
where $\epsilon_{ilm}$ are the components of the Levi-Civita tensor
density, $j_{mj}$ are the components of a dislocation flux density tensor
which govern the flow of Burgers vector in the medium and we have
adopted the convention that repeated indices are summed over. The dislocation
density tensor satisfies the relation,
\begin{equation}
\int \rho_{ij} dS_{i} = b_{j},
\end{equation}
where the integral extends over an arbitrary surface in the crystal, and
$b_{j}$ is the $j$-component of the sum of Burgers vectors of all the
dislocation lines that cross the surface.
It should also be noted here that the subsidiary constraint
\begin{equation}
\label{eq:solenoidal}
{\partial \rho_{ij} \over \partial x_{i}} = 0
\end{equation}
that the dislocation density be solenoidal must also be imposed to ensure
that dislocation lines do not terminate within the solid, except perhaps at
some internal interface such as a grain boundary.  The necessity of
incorporating this requirement was demonstrated by
Nabarro. \cite{re:nabarro52}

Equation (\ref{eq:conserve})
is correct microscopically, but involves the unknown flux $j_{mj}$,
now a second rank tensor. We note that this equation has already been used
as the starting point in other theories of dislocation dynamics
\cite{re:kosevich62,re:mura63} and that our approach differs in our
choice of the flux $j_{mj}$. As discussed earlier,
expressions for the flux are derived invoking the Peach-Koehler force
and continuum elasticity,
and therefore yield equations that are invariant under time reversal.
We suggest instead that dislocation motion
is overdamped at the coarse-grained scale of motion, and that a
{\rm constitutive} equation for the flux can be introduced
such that the dislocation density tensor evolves to minimize the local free
energy of the defected material. The origin of this dissipative motion lies
both in interactions with other microscopic degrees of freedom and
interactions with other dislocations within the coarse-graining
cell. As a consequence, our equations are manifestly not
invariant under time reversal.


As described earlier, the free energy of the crystal is a functional
of the dislocation density tensor, and a non-singular strain
$e_{ij}^{NS}$, which describes phonon modes.
It is useful to regard these two
tensor fields as part of the basis of an abstract Hilbert space of slow
variables which characterizes the system.  Further, if these two fields are
mutually orthogonal then, following Nelson and Toner \cite{re:nelson81},
\begin{equation}
\label{eq:free}
F  =  F_{NS}\left[e_{ij}^{NS}\right] + F_{d}\left[\rho_{ij}\right],
\end{equation}
where $F_{NS}$ and $F_{d}$ are the non-singular and the dislocation parts of
the free energy, respectively.

The spontaneous evolution of the system from some initial dislocation
configuration towards equilibrium results in a decrease in its free energy
(at constant temperature). We neglect in what follows the nonsingular
contribution to free energy dissipation since for the phenomena
of interest the associated time
scale for phonon propagation is expected to be much shorter than that for
dislocation motion. With this in mind, we have that the time rate of
change of the free energy is given by
\begin{equation}
\label{eq:prod}
\frac{d F}{dt} = \int d^{3}r \left({\delta F_{d} \over \delta
\rho_{ij}}\right ) \left({\partial \rho_{ij} \over \partial t}\right),
\end{equation}
where $\delta / \delta \rho_{ij}$ stands for functional or variational
derivative with respect to $\rho_{ij}$,
the integral extends over the entire system, and we have
chosen appropriate boundary conditions so that surface integrals vanish.
Upon substituting the conservation law (Eq. (\ref{eq:conserve})) and
integrating by parts, the requirements that the total free energy
be a monotonically decreasing function of time, and that there be a linear
relation between
forces and fluxes, leads to the identification of the dissipative
dislocation density flux,
\begin{equation}
\label{eq:dflux}
j_{mj} = - B_{mjst} \epsilon_{sab} {\partial \over \partial x_{b}} \left(
{\delta F_{D} \over \delta \rho_{at}} \right),
\end{equation}
where the fourth-rank tensor, $B_{mjst}$ plays the role of kinetic or
Onsager
coefficient, and embodies the complicated microscopic dynamics that
allows the reduction in free energy at the coarse-grained scale.
Substitution of Eq. (\ref{eq:dflux}) into the conservation law, Eq.
(\ref{eq:conserve}), leads to the dynamical equation for the
dislocation density tensor.

Further simplification is possible when the kinetic attributes of the
medium under consideration are spatially isotropic. In this case
the Curie principle \cite{re:groot84} constrains the form of
the tensor of kinetic coefficients to
\begin{equation}
\label{eq:onsager}
B_{mjst} = {\bar B} \left[{1 \over 2}\left(\delta_{mt} \delta_{js} +
\delta_{ms} \delta_{jt} \right) - {1 \over 3} \delta_{mj} \delta_{st}
\right],
\end{equation}
where $\bar B$ is the single required kinetic coefficient. It is also
possible to modify this description in order to distinguish between glide
and climb kinetics, though we will leave this modification for later work.
As our formulation focuses
on the coarse-grained dislocation density, the phenomenological constant
$\bar B$ can only provide limited insight into the details of the mechanism
for defect motion since, for example, the detailed geometry of slip planes
and directions has been averaged over.  Nevertheless, as will be shown
below, one can extract interesting microstructural information from our
model, and so there is an implicit information trade-off in this approach
to the problem.

The only quantity that remains to be specified is the free energy
density corresponding to the singular part of the deformation field,
and to express it explicitly as a function of the dislocation density tensor.
Kosevich \cite{re:kosevich79} has already calculated the elastic
energy of interaction between sets of dislocations, and this result was
later adopted by Nelson and Toner \cite{re:nelson81} for
equilibrium calculations \cite{fo:jeff1_1}.  The result is most conveniently
expressed in terms of an integration in reciprocal space by
\begin{equation}
\label{eq:fd}
F_{d} = {1 \over 2} \int {d^3{q} \over (2\pi)^{3}} K_{ijkl}(\vec{q})
\rho_{ij} (\vec {q}) \rho_{kl} (\vec {-q})
\end{equation}
with
\begin{equation}
K_{ijkl}(\vec{q})  =  K_{ijkl}^{0}(\vec{q}) +
K_{ijkl}^{c}(\vec{q}),
\end{equation}
where the main kernel $K_{ijkl}^{0}$ is given in terms of the shear modulus
$\mu$ and Poisson ration $\nu$ by
\begin{equation}
\label{eq:k0}
 K_{ijkl}^{0}(\vec{q}) = {\mu \over q^{2}} \left[Q_{ik} Q_{jl} + C_{il} C_{kj} +
{2\nu \over 1 - \nu} C_{ij} C_{kl} \right]
\end{equation}
and the core kernel $K_{ijkl}^{c}$ is given in terms of the edge and screw
core energies (per unit length), $E_{e}$ and $E_{s}$, by
\begin{equation}
\label{eq:kc}
 K_{ijkl}^{c}(\vec{q}) = 2E_{e} \delta_{ik} \delta_{jl} + 2(E_{s}-E_{e})
\delta_{ij} \delta_{kl}.
\end{equation}
It is advantageous to  express these kernels in terms of the transverse and
rotational $\vec{q}$-space projection operators
\begin{eqnarray}
\label{eq:transverse}
Q_{ij} = \delta_{ij} - {q_{i} q_{j} \over q^{2}},\\
C_{ij} = \epsilon_{ijl} {q_{l} \over q},
\end{eqnarray}
respectively, as will be evident below.

Before we proceed any further, we point out here
that, upon substituting the densities corresponding to single isolated
dislocations into Eq. (\ref{eq:fd}), one recovers the familiar
Peach-Koehler interaction \cite{re:hirth68}.  For example, consider two
infinite, parallel screw dislocations, with Burgers vectors $\vec{b}^{(1)}$
and $\vec{b} ^{(2)}$, which are located at the origin and at $\vec{a}$,
respectively. The Fourier transforms of the dislocation densities
corresponding to these defects are then
\begin{eqnarray}
\label{eq:ddens}
\rho_{ij}^{(1)}(\vec{q}) = 2\pi\; b^{(1)} \delta_{i3}\; \delta_{j3}\;
\delta(q_{z}),\\
\rho_{ij}^{(2)}(\vec{q}) = 2\pi\; b^{(2)}  \delta_{i3} \;\delta
_{j3} \;\delta(q_{z}) e^{-i\vec{q} \cdot \vec{a}}
\end{eqnarray}
and the line direction is along the $z$-axis. Neglecting the core
contributions for the
moment, one finds that the interaction free energy per unit dislocation
length is
\begin{equation}
\label{screwfree}
- {\mu b^{(1)} b^{(2)} \over 2\pi^{2}} \int {d^{2} \; q {1 \over q^{2}}\;
e^{-i \vec{q} \cdot \vec{a}} }  =  - {\mu \over 2\pi} b^{(1)} b^{(2)} \ln(a),
\end{equation}
implying that the associated force (per unit length) =
$\mu b^{(1)} b^{(2)}/2\pi a$, as expected.
Thus, the main kernel
given by Eq. (\ref{eq:k0}) embodies long-ranged interactions among
dislocations.

The effect of an external stress field with components $\sigma_{ij}^{0}$ on
a collection of dislocations can also be incorporated by the addition to
the free energy $F_{d}$, of a term representing the interaction of the
dislocation density with the external field.
Following Mura \cite{re:mura87}, we write $F_{ext}$ as
\begin{equation}
\label{eq:external}
F_{ext} = - \int d^{3} r \;\sigma_{ij}^{0} \; \epsilon_{ij}^{*},
\end{equation}
where $\epsilon_{ij}^{*}$ is the eigenstrain associated with a particular
dislocation density tensor configuration,
and the total free energy is then $F_{tot}$ = $F_{d}$ + $F_{ext}$.
We will elaborate further on this issue below.

By combining the results summarized above, one finally arrives at the
equation of motion for the dislocation density tensor,
\begin{equation}
\label{eq:eqofmot}
{\partial \rho_{ij}(\vec{q}) \over \partial t} = q^{2} C_{mi} C_{sa}
B_{mjst}{\delta F_{tot} \over \delta \rho_{at}(-\vec{q})}.
\end{equation}

\section{Stationary solutions}
\label{sec:stationary}

Clearly, equation (\ref{eq:eqofmot}) is linear in the dislocation density
tensor
and will therefore only describe the linear relaxation of a defected system
towards the asymptotic state of thermodynamic equilibrium, given by
the minimization of $F_{tot}$. The relaxation rate as a function of
wavevector $\vec{q}$ can be directly obtained from the eigenvalues of the
kernel of Eq. (\ref{eq:eqofmot}).

It is of interest to first study
those eigenmodes of zero eigenvalue; i.e., the stationary solutions.
We start by examining the equilibrium solutions, i.e., configurations of
defects that minimize the free energy.

\subsection{Equilibrium configuration without external stress}

We first consider the free energy $F_{d}$ in the absence of external
stresses and seek those densities that satisfy
\begin{equation}
\label{eq:gstate}
{\delta F_{d} \over \delta \rho_{at}(-\vec{q})} = \left[ K_{atpq}^{0}(\vec{q})
+ K_{atpq}^{c}(\vec{q}) \right] \rho_{pq}(\vec{q}) = 0.
\end{equation}
The kernel can be regarded as a 9x9 Voigt matrix with components
$K_{\alpha \beta}$ ($1 \leq \alpha, \beta \leq 9$) and, since
$K_{\alpha \beta}$
=  $K_{\beta \alpha}$, its eigenvalue spectrum is real.  This kernel is
non-singular and therefore the only eigenstates with zero
eigenvalue are the trivial ones: $\rho_{pq}=0$, i.e., a defect free medium.
This fact inmediately follows from the core contribution to the
free energy (Eq. (\ref{eq:kc})).  However, Eq. (\ref{eq:gstate}) also
supports defected equilibrium states for a system constrained to
having nonzero densities $\rho_{pq}$. The Lagrange multipliers for
such a constrained minimization can be chosen so as to cancel the
core contribution, so that the problem
therefore reduces to finding the (non-trivial) eigenstates of the
kernel $K_{atpq}^{0}$  which have zero eigenvalue.   This latter kernel is
singular,
hence one expects that there will be at least one such zero eigenstate.

For this purpose it is convenient to express the dislocation density in
terms of the operators which comprise the main kernel. For
a rotationally invariant system one can form
three second-rank tensors from $\vec{q}$: $\delta_{ij}$, $q_{i} q_{j}$ and
$\epsilon_{ijk} q_{k}$. We therefore use the decomposition
\cite{re:forster75}
\begin{equation}
\label{eq:rhoexp}
\rho_{ij}(\vec{q}) = \delta_{ij} X(q) + P_{ij} Y(q) + C_{ij} Z(q),
\end{equation}
where the longitudinal projector $P_{ij} = \delta_{ij} - Q_{ij}$, and where
$X$, $Y$ and $Z$ are three $q$-dependent amplitudes.  Such a decomposition
has the advantage that it permits the classification of the density in
terms of its parity since, for example, $P_{ij}$ and $C_{ij}$ have
different signatures under parity.  Also, it is possible to determine the
edge and screw character of the density by noting that for an edge (screw)
dislocation the Burgers vector $\vec{b}$ is perpendicular (parallel or
anti-parallel) to its line direction.  Hence, $\rho_{ij}$ is of edge (screw)
character if $\rho_{ij} \neq 0$ for $i \neq j \; (i = j)$.
%
%
The effect of the main kernel on the
dislocation density can then be determined from the results of the operator
algebra summarized below:
\begin{equation}
\label{eq:algebra1}
C_{sa} C_{aq} = -Q_{sq}
\end{equation}
\begin{equation}
\label{eq:algebra2}
C_{sa} Q_{ap} = C_{sp} = Q_{sa} C_{ap}
\end{equation}
\begin{equation}
\label{eq:algebra3}
Q_{ia} Q_{ab} = Q_{ib}
\end{equation}

With the preceding relations is mind, one can immediately identify several
degenerate eigenstates with zero eigenvalue by observing that
\begin{eqnarray}
\label{eq:eigenstate1}
K_{atpq}^{0} Q_{pq} = 0,\\
\label{eq:eigenstate2}
K_{atpq}^{0} \delta_{pq} = 0,
\end{eqnarray}
and, consequently, $P_{pq}$ is also an eigenstate.
For the problem under
consideration, then, $Q_{pq}$ is an acceptable eigenstate as it is solenoidal
while $\delta_{pq}$ clearly does not satisfy this requirement. In the
former case, the
transverse operator can be represented in real space by performing an
inverse Fourier transformation to obtain
\begin{equation}
\label{eq:qpq}
Q_{pq}(\vec{r})  = {1 \over (2\pi)^{3}} \left[ \left ({1
\over r^{3}} \delta_{pq} - {3 x_{p} x_{q} \over r^{5}} \right) + 2
\delta(\vec{r}) \delta_{pq} \right]
\end{equation}
Thus, the dislocation density falls off like 1/$r^{3}$ and has a
quadrupolar angular dependence.

There are other degenerate eigenstates which
correspond to more spatially localized dislocation densities and that
break rotational invariance.  As an
example, consider the line along $q_{1}$ in reciprocal space
\begin{equation}
\label{eq:qline}
\rho_{pq} = A \left [ \delta_{pq} - \delta_{p1} \delta_{q1} \right]
\delta(\vec{q_{2}}) \delta(\vec{q_{3}}),
\end{equation}
which is an eigenstate of the main kernel with zero eigenvalue, given that A
is a dimensional constant that is independent of wavevector.  This
particular combination of delta functions was chosen to ensure that the
density is solenoidal.  Now, this line in reciprocal space maps into the
$x_{2}-x_{3}$ plane in real space.  Further, the fact that the only two
non-zero components are $\rho_{22}$ and $\rho_{33}$ implies that the
density has screw character.  In short, Eq. (\ref{eq:qline}) corresponds to
a two-dimensional dislocation \lq\lq wall" in real space.  The screw
character of the wall implies that it can be regarded, in some sense, as a
prototypical (low-angle) twist grain boundary.  It is
interesting to
note that such a regular distribution of dislocations might be expected
from the tendency of dislocations to polygonize.  In this model, however,
such a structure has been obtained without postulating the existence of
additional \lq\lq chemical" forces, related to vacancy concentration at
dislocations, which inhibit climb.  Walls of other orientations are, of
course, also possible.

\subsection{With External Stress}

In many cases of interest the defected system will be subjected to an
externally imposed stress field. Equation (\ref{eq:eqofmot}) also
applies in this case, with the free energy explicitly including the
contribution arising from the imposed stress.
Now, in order to apply our previous results, it is first necessary to relate
the eigenstrain in Eq. (\ref{eq:external}) to the dislocation density.
Such a relation may be constructed by noting that the eigenstrains for this
problem arise from plastic distortion, and that the essential connection
between plastic strain and the Burgers vector of a dislocation can be
expressed in terms of the closure failure of a Burgers circuit that
encircles the dislocation. Generalizing this idea to a collection of
dislocations leads to
\begin{equation}
\label{eq:circuit}
\rho_{ij}(\vec{r}) =  - \epsilon_{ilm} {\partial \beta^{P}_{mj} \over
\partial x_{l}},
\end{equation}
where the components of the plastic distortion tensor $\beta^{P}_{ij}$ are
related to the eigenstrain by $\epsilon_{ij}^{*} = (1/2) (\beta^{P}_{ij}
+ \beta^{P}_{ji})$.  It should be noted here that Eq. (\ref{eq:circuit}) by
itself does not uniquely define the plastic strain.  In particular, it does
not specify the divergence of the plastic strain and so some additional
input, such as the specification of a gauge, is required. The role of
gauge invariance in the field theory of dislocations has been discussed
elsewhere. \cite{re:golebiewska79}.
In the present context this additional input is
contained implicitly in Eq. (\ref{eq:dflux}) since
the dislocation flux is related to the time-rate change of the plastic
distortion by $j_{mk} = - \partial \epsilon_{mk}^{*}/\partial t$, a relation
indicated by Kroner and Rieder \cite{re:kroner56}.  Further, from the form
on the kinetic coefficient employed in Eq. (\ref{eq:dflux}), it is evident
that $j_{mm} = 0$, and so there is no time rate change of the volume
associated with plastic deformation.

With this in mind, we consider here, for the purpose of illustration, the
specific case of a symmetric, solenoidal  external stress tensor given in
reciprocal space by
\begin{equation}
\label{eq:ext}
\sigma_{ij}^{0}(\vec{q}) = \bar{\sigma}(q) Q_{ij},
\end{equation}
where $\bar{\sigma}(q)$ is a function of the wavenumber $q$. For this case
$F_{ext}$ becomes
\begin{equation}
\label{eq:fext1}
F_{ext} = - \int {d^{3} q \over 2\pi^{3}}\; \bar{\sigma}(q) Q_{ij}
\epsilon_{ij}^{*}(-\vec{q})
=  i \int {d^{3} q \over 2\pi^{3}} \; { \bar{\sigma}(q) \over q}  C_{ij}
\rho_{ij}(-\vec{q}),
\end{equation}
where Eq. (\ref{eq:circuit}) has been used to relate the transverse
projection of the dislocation density to the rotational projection.  So,
the state of equilibrium is defined by the eigenvalue equation
\begin{equation}
\label{eq:gst1}
{\delta F_{tot} \over \delta \rho_{at}(-\vec{q})} = K_{atpq}^{0}(\vec{q})
\rho_{pq
}(\vec{q}) + {i \over q} \bar{\sigma} C_{at} = 0.
\end{equation}
A solution to this eigenvalue problem can be obtained by noting that
$C_{ij}$ itself is an eigenstate of the main kernel since
\begin{equation}
\label{eq:ceig}
K_{ijkl}^{0} C_{kl} = 2 {1 + \nu \over 1 - \nu} C_{ij},
\end{equation}
and so an equilibrium configuration (in reciprocal space) is given by
\begin{equation}
\label{eq:est}
\rho_{pq}(\vec{q}) = {-i \over 2 q} C_{pq} \bar{\sigma}(q) {1 - \nu \over 1
+ \nu},
\end{equation}
or in real space by
\begin{equation}
\label{eq:rhor}
\rho_{pq}(\vec{r})  =  \int d^{3} r' \bar{\sigma}(\vec{r}^{~ \prime})
L_{pq}(\vec{r}-\vec{r}^{~ \prime}),
\end{equation}
where
\begin{equation}
\label{eq:lpq}
L_{pq}(\vec{r}) = {- 1 \over 8 \pi}  {1 - \nu \over 1 + \nu} {\epsilon_{pql}
x_{l} \over r^{3}}.
\end{equation}
Thus, the application of a \lq\lq transverse" stress field results in a
dislocation density with \lq\lq rotational" character.

Since, as indicated above, Eq. (\ref{eq:dflux}) relates the time-rate
change of the eigenstrain to the dislocation density, it is advisable in
the case of an arbitrary (symmetric) external stress to first develop an
equation for the temporal evolution of the Fourier transform of the
eigenstrain, $\epsilon^{*}_{ij}(\vec{q})$, and then determine the
dislocation density by using the Fourier transform of
Eq. (\ref{eq:circuit})
\begin{equation}
\label{eq:circuitf}
\rho_{ij}(\vec{q}) = - i q C_{im} \beta^{P}_{mj}.
\end{equation}
or the symmetrized form
\begin{equation}
\label{eq:symm}
q^{2} \;\epsilon_{ilm} \epsilon_{jpk} \;P_{lp}\epsilon^{*}_{mk} =  \eta_{ij},
\end{equation}
where $\eta_{ij}$ is Kroner's incompatibility tensor \cite{re:kroner58},
which is essentially the Ricci tensor and therefore a measure of
curvature. \cite{re:weinberg72}
Such an equation can be determined by using Equations (\ref{eq:dflux}),
(\ref{eq:k0}),
(\ref{eq:gstate}) and (\ref{eq:external}) to obtain
\begin{equation}
\label{eq:plasticeq}
{\partial \epsilon^{*}_{mk} \over \partial t} =  \mu B_{mkst}
\left[Q_{sq} Q_{tp} + Q_{sp} Q_{tq} -
{2\nu \over 1 - \nu} Q_{st} Q_{pq} - \sigma^{0}_{pq} \right] \epsilon^{*}_{pq}.
\end{equation}
Thus, both the dislocation density and the eigenstrain can be determined
from this formulation.  Further, since the dependence of the eigenstrain
on time is dependent on the dislocation flux, the value of the
eigenstrain will depend upon the time history of the system, as expected.
%
%

Equations (\ref{eq:circuitf}) and (\ref{eq:plasticeq}) can be used to
identify the dislocation density corresponding to a specific plastic
strain. For example, in the case of no externally applied stress field, the
plastic distortion
$\beta^{P}_{ij} = P_{ij}$ corresponds to $\rho_{ij} = 0$,
whereas the plastic distortion $\beta^{P}_{ij} = C_{ij}$ (and therefore
$\epsilon^{P}_{ij}$ =$0$) corresponds to  $\rho_{ij} = Q_{ij}$. In each
case, the plastic strain and the corresponding dislocation density of these
equilibrium states are time-independent.

Finally we address the issue of whether there exist other stationary
solutions of the dynamic equations that are not minima of the free
energy. We show next that the tensor $C_{mi}C_{sa}B_{mjst}$, the
generalized kinetic coefficient in Eq. (\ref{eq:eqofmot}),
commutes with the linear kernel $K^{0}$ for the class of
solutions that do not break rotational invariance. Hence all the
stationary solutions of the dynamics are minima of the free energy.
Given the general decomposition for a rotationally invariant system
of Eq. (\ref{eq:rhoexp}), and the fact that $\delta_{ij}$, $Q_{ij}$
and $C_{ij}$ are eigenstates of $K^{0}$, it is sufficient to show
that $\delta_{ij}$, $Q_{ij}$ and $C_{ij}$ are also eigenstates of
$C_{mi}C_{sa}B_{mjst}$.

First, by using Eqs. (\ref{eq:onsager}) and (\ref{eq:algebra2}), we
find
\begin{equation}
C_{mi}C_{sa}B_{mjst} Q_{at} = \bar{B} C_{mi} \left[ \frac{1}{2}
\left( C_{jm}
+ C_{mj} \right) - \frac{1}{3} \delta_{mj} C_{ss} \right] = 0.
\end{equation}
The first term vanishes because $C_{ij}$ is an antisymmetric tensor. The
second vanishes because $C_{ij}$ is proportional to the Levi-Civita
tensor. Similarly, by using Eqs. (\ref{eq:onsager}),(\ref{eq:algebra1})
and (\ref{eq:algebra2}) we find
\begin{equation}
C_{mi}C_{sa}B_{mjst}C_{at} = - \bar{B} C_{mi} \left[
\frac{1}{2} \left( Q_{mj} + Q_{jm} \right) - \frac{1}{3} \delta_{mj}
Q_{ss} \right] = \bar{B} \left( - C_{ij} + \frac{2}{3} C_{ij}
\right),
\end{equation}
where we have used the fact that $Q_{ij}$ is symmetric and that
$Q_{ss} = 2$.
Finally,
\begin{equation}
C_{mi}C_{sa}B_{mjst} \delta_{at} = \bar{B} C_{mi} \left[
\frac{1}{2} \left( C_{mj} + C_{jm} \right) - \frac{1}{3} \delta_{mj}
\delta_{ts} C_{st} \right] = 0.
\end{equation}

\section{Conclusions}
\label{sec:conclusion}

We have considered a formal coarse-graining procedure over
elements of volume that contain a
large number of dislocation lines, and argued that microscopic
interactions among the lines within each element of volume give rise to
dissipative macroscopic dynamics. A constitutive law relating the
dislocation density tensor and its associated flux has been
introduced such that the evolution of the defected system is driven by the
minimization of its plastic free
energy. A standard procedure based on linear irreversible
thermodynamics leads to the equations of motion for the dislocation
density.

For the case of an isotropic medium, the energy of interaction between
dislocations in terms of the dislocation density tensor
had already been calculated by Kosevich. We have further
shown that in this case the tensor of kinetic coefficients that appears in
the kinetic law requires only the
specification of a scalar quantity, and that it commutes with the kernel of
the linear operator driving energy minimization. Some stationary solutions
have been explicitly constructed, both in systems with and without
externally imposed stresses.

We have restricted our attention in this paper to idealized
systems wherein there is no mechanism for dislocation production, such as a
Frank-Read source. \cite{re:hirth68}
It is possible, however, to incorporate a stochastic
\lq\lq source" term in Eq. (\ref{eq:dflux}) satisfying a
fluctuation-dissipation relation to model dislocation production.
This would be useful in examining, for example, the change in mechanical
properties that attend dislocation formation, such as work hardening.

Finally, given that the approach discussed here is based on
having
partially averaged microscopic degrees of freedom, it seems of interest
to relate it to results from microscopic descriptions such as
the ones outlined in the
Introduction. For example, one might consider the dynamical evolution
that results from these other models and then
invoke a coarse-graining procedure.
This, of course, would require systems that contain a large number
of dislocations and, as mentioned above, algorithms for such large
systems have or are currently being developed.  Thus, it may be
possible to bridge the disparate length scales between both
descriptions.

\section*{Acknowledgments}
We are indebted to Y.T. Chou, Hamid Garmestani, W. W. Mullins and David
Srolovitz for useful
discussions. The work of JMR is supported by
the National Science Foundation under contract DMR-9458028.
JV is supported by the U.S. Department of Energy, contract No.
DE-FG05-95ER14566, and also in part by the Supercomputer
Computations Research Institute, which is partially funded by the U.S.
Department of Energy, contract No. DE-FC05-85ER25000.

\bibliographystyle{prsty}
\bibliography{references}

\begin{thebibliography}{10}

\bibitem{re:hull92}
D. Hull and D. Bacon, {\em Introduction to Dislocations} (Pergamon Press, New
  York, 1992).

\bibitem{re:hirth68}
J. Hirth and J. Lothe, {\em Theory of Dislocations} (McGraw-Hill, New York,
  1968).

\bibitem{re:mura63}
T. Mura, Phil. Mag {\bf 8},  843  (1963).

\bibitem{re:ghonhiem88}
N. Ghonhiem and R. Amodeo,  in {\em Nonlinear Phenomena in Materials Science},
  edited by G. Morton and L. Kubin (Transtech Publications, Aedermannsdorf,
  Switzerland, 1988), p.\ 377.

\bibitem{re:gulluoglu90}
A. Gulluoglu, D. Srolovitz, R. LeSar, and P.~S. Lomdahl,  in {\em Simulation
  and Theory of Evolving Microstructures} (TMS, Warrendale, PA, 1990).

\bibitem{re:mura87}
T. Mura, {\em Micromechanics of Defects in Solids} (Martinus Nijhoff, Boston,
  1987).

\bibitem{re:gulluoglu89}
A. Gulluoglu, D. Srolovitz, R. LeSar, and P. Lomdahl, Scripta Met. {\bf 23},
  1347  (1989).

\bibitem{re:turner82}
A. Turner and B. Hasegawa, ASTM  761  (1982).

\bibitem{re:nelson81}
D. Nelson and J. Toner, Phys. Rev. B {\bf 24},  363  (1981).

\bibitem{re:kosevich79}
A. Kosevich,  in {\em Dislocations in Solids}, edited by F. Nabarro
  (North-Holland, New York, 1979), Vol.~1.

\bibitem{re:brand84}
H. Brand and K. Kawasaki, J. Phys. A {\bf 17},  L905  (1984).

\bibitem{re:kawasaki84a}
K. Kawasaki, Prog. Theor. Phys. Suppl. {\bf 79},  161  (1984).

\bibitem{re:kawasaki84b}
K. Kawasaki, Prog. Theor. Phys. Suppl. {\bf 80},  123  (1984).

\bibitem{re:cross93}
M. Cross and P. Hohenberg, Rev. Mod. Phys. {\bf 65},  851  (1993).

\bibitem{re:martin72}
P. Martin, O. Parodi, and P. Pershan, Phys. Rev. A {\bf 6},  2401  (1972).

\bibitem{re:forster75}
D. Forster, {\em Hydrodynamic Fluctuations, Broken Symmetry, and Correlation
  Functions} (Bejamin/Cummings, Reading, MA, 1975).

\bibitem{re:nabarro52}
F. Nabarro, Adv. in Phys. {\bf 1},  284  (1952).

\bibitem{re:kosevich62}
A. Kosevich, Soviet Phys. JEPT {\bf 15},  108  (1962).

\bibitem{re:groot84}
S. de~Groot and P. Mazur, {\em Non-equilibrium Thermodynamics} (Dover, New
  York, 1984).

\bibitem{fo:jeff1_1}
In order to simplify the notation somewhat, we ignore the notational
  distinction between a function and its Fourier transform and, in this case,
  represent each by the same tensor field $\rho_{ij}$.

\bibitem{re:golebiewska79}
A. Golebiewska-Lasota, Int. J. Engng. Sci. {\bf 19},  329  (1979).

\bibitem{re:kroner56}
E. Kroner and G. Rieder, Zs. Phys. {\bf 145},  424  (1956).

\bibitem{re:kroner58}
E. Kroner, {\em Kontinuumstheorie der Versetzungen and Eigenspannugen}
  (Springer, Berlin, 1958).

\bibitem{re:weinberg72}
S. Weinberg, {\em Gravitation and Cosmology. Principles and Applications of the
  General Theory of Relativity} (J. Wiley \& Sons, New York, 1972).

\end{thebibliography}

\end{document}